# Quantitative analysis of iron sand mineral content from the south coast of Cidaun, West Java using rietveld refinement method

Novianty Rizky Ardiani, Setianto Setianto, Budy Santosa, Bambang Mukti Wibawa, Camellia Panatarani, and I Made Joni

View Online

Export Citation

## ARTICLES YOU MAY BE INTERESTED IN

Physicochemical characterizations and insecticidal properties of Lantana camara leaf ethanolic extract with powder application
AIP Conference Proceedings **2219**, 040002 (2020); https://doi.org/10.1063/5.0003200

The effectiveness of suspension of Beauveria bassiana mixed with silica nanoparticles (NPs.) and carbon fiber in controlling Spodoptera litura
AIP Conference Proceedings **2219**, 080011 (2020); https://doi.org/10.1063/5.0003159

Preparation of precipitated calcium carbonate from carbon mineralization of raw biogas with Ca(OH)$_2$ solution using bubble column contactor
AIP Conference Proceedings **2219**, 030004 (2020); https://doi.org/10.1063/5.0003060

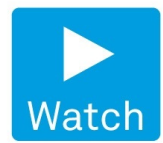
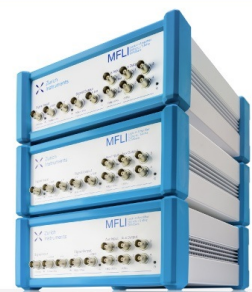

# Quantitative Analysis of Iron Sand Mineral Content from The South Coast of Cidaun, West Java using Rietveld Refinement Method

Novianty Rizky Ardiani[1], Setianto Setianto[1, a)], Budy Santosa[2], Bambang Mukti Wibawa[3], Camellia Panatarani[1, 4] and I Made Joni[1, 4]

[1]*Department of Physics, Faculty of Mathematics and Natural Sciences, Universitas Padjadjaran,*
[2]*Department of Geophysics, Faculty of Mathematics and Natural Sciences, Universitas Padjadjaran,*
[3]*Department of Electrical Engineering, Faculty of Mathematics and Natural Sciences, Universitas Padjadjaran,*
[4]*Nanotechnology and Graphene Research Centre, Universitas Padjadjaran,*
*Jl. Raya Bandung-Sumedang KM 21, Jatinangor, West Java 45363, Indonesia.*

a)Corresponding author : setianto@phys.unpad.ac.id

**Abstract**. Iron sand is one of the abundant natural resources in Indonesia, especially on the south coast of Cidaun; West Java which is the basic material for building and metal industry. Iron mineral content is generally metal oxide such as magnetite, hematite and silica/quartz. Sand with iron content used in this study is derived from beach sand Desa Kertajadi, Kecamatan Cidaun, Kabupaten Cianjur, Jawa Barat. Then mass of 2 kg sand was separated using a magnetic separator in order to obtain magnetic and nonmagnetic mineral content. After nine rounds of separation takes two different types of samples that are no separation sand (TS) sample and concentrate in the third separation (S3) sample. The sample is then examined by X-Ray Diffraction (XRD) measurement and analyzed quantitatively using MAUD software to determine the content of $Fe_3O_4$ (magnetite) by using the Rietveld refinement method from XRD data. As the analysis result, the magnetite content contained in iron sand is counted quantitatively for each different sample. For iron sand samples (TS) yielding a 24.27 percent of magnetite and a third concentrate separation sample (S3) yields 61.98 percent.

## INTRODUCTION

Sand on the beach is usually only used for building materials. In fact, sand has a high potential besides its use as a construction material, because it was found that beach sand contains a lot of minerals. The mineral content is divided into three groups, namely strong magnetic minerals, weak magnetic minerals and non-magnetic minerals [1]. The highly magnetic mineral content contained in the sand is usually present in the form of iron ore $Fe_3O_4$ (magnetite). The content of weakly magnetic minerals contained in the sand is usually $Fe_2O_3$ (hematite). Non-magnetic minerals contained in the sand are $SiO_2$ (silica). In iron sand, iron content in the form of iron oxide, iron oxide, which is usually contained in iron sand, is magnetite and hematite. Magnetite has a strong interaction with magnets compared to hematite. Magnetite is a strongly magnetic mineral, has a cubic spinel crystal structure. Magnetite is also contained in the ferrimagnetic properties at room temperature [2]. Magnetite is a mineral with a bandgap of 0.1 eV, black color and conductivity of $10^2$-$10^3$ Ω/cm [3]. Magnetite is indispensable for industrial development which can be used as a mixture of colors, basic components of dry ink (toner) on copiers and laser printers [4]. In addition, magnetite can be used as a basis for the production of permanent magnets because of the strong reaction of magnetite to external magnets. Permanent magnets are needed by the industry as an energy source and are very useful in various electronic device components. From the results of previous studies based on the results of X-ray diffraction characterization (XRD), it was found that magnetic minerals magnetite, hematite and (Fe, Mg) (Cr, Fe)$_2$O$_4$ contained in the iron sand of the Kebumen area at the Ambal beach (magnetite iron) were magnesium chromium oxide) [5]. In addition, the XRD test results on the south coast of Bantul Yogyakarta are magnetic materials dominated by mineral magnetite and maghemite [6]. While iron sand from Tegal Buleud, Sukabumi's south coast, magnetite, ilmenite, and hematite-titano, non-magnetic minerals contained $SiO_2$, $CaO$, $MgO$, $Cr_2O_3$, and $Al_2O_3$ [7]. Therefore, an investigation is needed to determine the mineral content of the sand. To determine the mineral content contained in beach sand, it can be analyzed quantitatively by taking samples that are processed with magnetic separators. This is done to separate magnetic minerals and non-magnetic minerals; the process is carried out several times to get several samples of iron sand. Samples were then crushed and filtered using 200 mesh. Subsequently, samples were characterized using X-ray diffraction (XRD). We used MAUD software to determine the quantitative results of the magnetite contained in the sand.

## MATERIALS AND METHOD

In general, for the sample preparation process, iron sand from mining products on the south coast of West Java is taken in varied portions according to the location. Each sample in the form of powder/granules is sieved by using a sieve of 60 mesh up to 200 mesh. The sand drying is carried out to eliminate the water content contained in the sand. This sand drying will be helpful during the separation process because the magnetic separator used is a magnetic separator type dry drum separator. The process for drying sand is carried out in the oven using a drying oven at a temperature of 100 ° C. The results of the drying are then taken up to 10 grams, which then becomes a sample of non-separating sand (TS). The separation is performed to separate magnetic minerals and non-magnetic minerals. The sand removal takes place with a magnetic separator. The type of magnetic separator used is a dry magnetic separator of low intensity. The separation for 2 kg takes place in 9 stages, with each stage being weighed. For the $1^{st}$ separation cut test (SS1) and the $3^{rd}$ concentrate sample (S3), the sample was taken 10 grams for the XRD test. The very wide sieving is used for separating the size of 300 mm to a size of about 40 μm, although efficiency decreases with the smoothness [8-10]. XRD measurements were then performed to determine the crystal structure of the iron sand samples. The intensity pattern plot obtained from the XRD results is analyzed quantitatively using the Material Analysis Using Diffraction (MAUD) application. The magnetite content is determined by matching the XRD pattern graph to a database reference from the Crystallography Open Database (COD). Thereafter, the data are obtained in the form of quantitative results for magnetite minerals. The database used is magnetite, hematite, graphite, silica alpha and clinoferrosilite. For this test, the selected samples are TS and S3. For the three-fold separation, a sample was chosen because, from the separation results, the color of the observed separation did not deviate significantly from the results of two-fold and four-fold separation.

## RESULTS AND DISCUSSION

### Patterns Analysis for Iron Sand Samples (TS)

The intensity pattern of the iron sand sample was obtained from XRD measurement before the separator process is shown in Fig. 1. The highest intensity peak is at position 27 degrees with which is exactly of the silicon oxide-alpha database. Certainly, the most minerals contained in the TS samples are silicon oxide-alpha. We found carbon content (graphite-like) at position 25.5 degrees. The iron ore minerals like magnetite, clinoferrosilite and hematite are at the position around 35 degrees. The quantitative analysis results of TS samples are generated Rietveld refinement method using MAUD software can be seen in Table 1. From Table 1, the highest minerals contained in TS sample were silicon oxide-alpha with 61.77 percent. The second highest content is a magnetite of 24.27 percent. Carbon content is about 8.34 percent. Also, there is a clinoferrosilite of 4.73 percent and the least is the hematite of 0.89 percent.

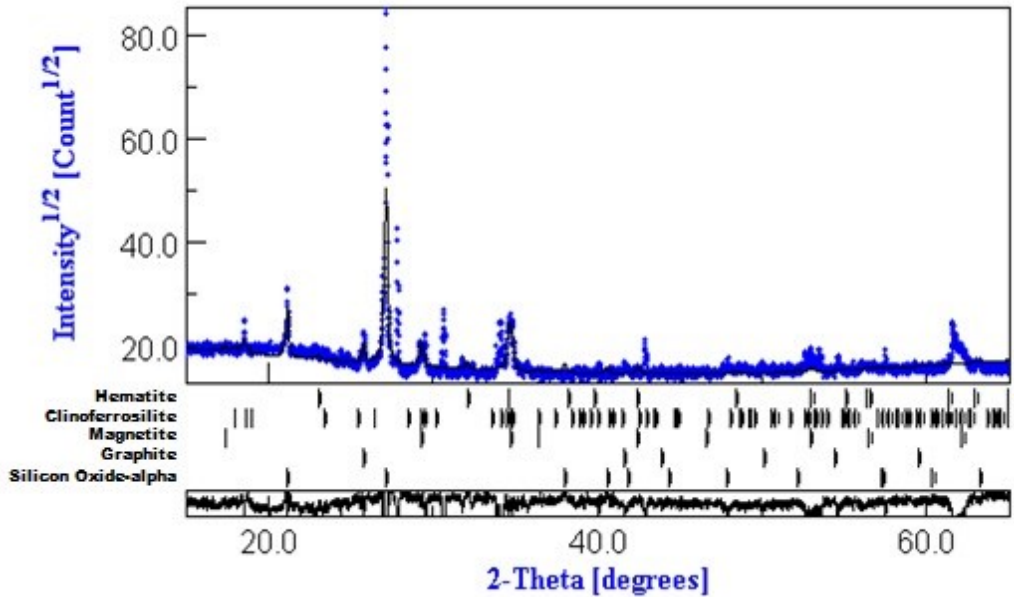

**FIGURE 1.** XRD pattern and quantitative analysis results of TS samples

**TABLE 1.** Quantitative results of TS samples

| Mineral | Quantitative (%) |
|---|---|
| *Magnetite* ($Fe_3O_4$) | 24.27 |
| *Hematite* ($Fe_2O_3$) | 0.89 |
| *Silicon oxide-alpha* ($SiO_2$) | 61.77 |
| Carbon (C) | 8.34 |
| *Clinoferrosilite* ($FeSiO_3$) | 4.73 |

## Pattern Analysis of Iron Sand Concentrate from Third Separation (S3)

The XRD pattern of iron sand concentrate (S3) samples is shown in Fig. 2. Intensity pattern for S3 sample shows that the highest intensity peak is at position 35 degrees. The minerals content on this peak is constructive interference of magnetite, hematite, and clinoferrosilite. The quantitative results from S3 samples can be seen in Table 2.

As a result, the highest minerals contained in the S3 sample are magnetite of 61.98 percent. The second highest content is an amorphous Carbon of 30.03 percent. The third highest content was silicon oxide-alpha of 4.92 percent. Then, there is clinoferrosilite of 3.00 percent and the least is hematite only 0.07 percent. Comparing the quantitative results, the magnetite content of S3 is higher than TS sample. This quantitative value for the magnetite content of the S3 sample shows that the separation process with the magnetic separator is available. The magnetic separator used successfully manages the impurities contained in the iron sand so that the magnetite mineral becomes more recovery. The quantitative results for hematite in S3 samples were smaller when compared with TS samples. This small amount of hematite mineral in S3 sample occurs because the hematite is a weak magnetic material so that when the separation process occurs the hematite will fall caused by the force of gravity. In other words, the hematite property is a weak magnetic whose magnetic force is not large enough to adhere to the drum separator. The quantitative results for carbon content in S3 samples are greater when compared with TS samples. It shows during the process of separation, much carbon attached to the drum separator. Clinoferrosilite content was less than 5 percent both S3 and TS samples. This suggests that the separator yield by using a magnetic separator may not reduce clinoferrosilite. Last, the silicon oxide-alpha content in S3 samples was fewer compared to TS samples. It suggests that the separation process using magnetic separators can reduce the silicon oxide-alpha in the sand successfully.

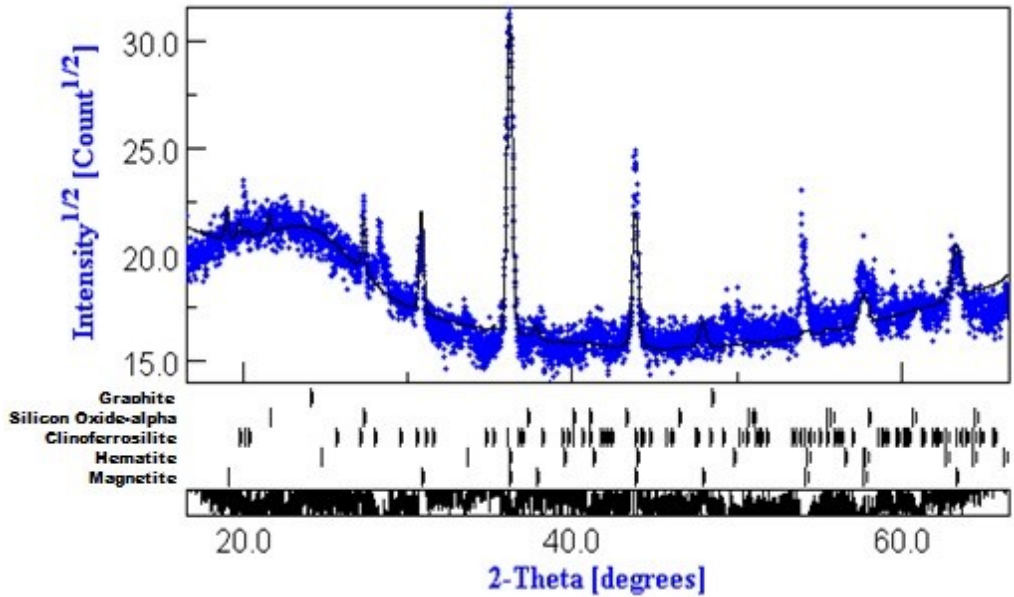

**FIGURE 2.** XRD Pattern and quantitative analysis results of S3 sample

**TABLE 2.** the quantitative results of S3 samples

| Mineral | Quantitative (%) |
|---|---|
| *Magnetite* ($Fe_3O_4$) | 61.98 |
| *Hematite* ($Fe_2O_3$) | 0.07 |
| *Silicon oxide-alpha* ($SiO_2$) | 4.92 |
| C (amorphous Carbon) | 30.03 |
| *Clinoferrosilite*($FeSiO_3$) | 3.00 |

## CONCLUSION

Finally, based on quantitative analysis results using the Rietveld refinement method, we can determine the magnetite mineral content in iron sand for each sample effectively. For iron sand ore samples (TS) yield 24.27 percent of magnetite content and 61.98 percent for concentrate sample (S3) from the third separation process. In the future, we should have calculated the recovery of magnetite content in iron sand via this process.

## ACKNOWLEDGMENT

This work is supported by Hibah Internal Unpad (HIU-2018) for Fundamental Research Project under contract number 476 d/UN6.RKT/LT/2018.